# Asynchronous Remote Medical Consultation for Ghana

Rowena Luk
Intel Research
2150 Shattuck Ave., Ste. 1300
Berkeley, CA 94704-1347  USA
rowenaluk@gmail.com

Melissa Ho
School of Information
Univ. of California, Berkeley
Berkeley, CA 94720-4600  USA
mho@ischool.berkeley.edu

Paul M. Aoki
Intel Research
2150 Shattuck Ave., Ste. 1300
Berkeley, CA 94704-1347  USA
aoki@acm.org

**ABSTRACT**
Computer-mediated communication systems can be used to bridge the gap between doctors in underserved regions with local shortages of medical expertise and medical specialists worldwide. To this end, we describe the design of a prototype remote consultation system intended to provide the social, institutional and infrastructural context for sustained, self-organizing growth of a globally-distributed Ghanaian medical community. The design is grounded in an iterative design process that included two rounds of extended design fieldwork throughout Ghana and draws on three key design principles (social networks as a framework on which to build incentives within a self-organizing network; optional and incremental integration with existing referral mechanisms; and a weakly-connected, distributed architecture that allows for a highly interactive, responsive system despite failures in connectivity). We discuss initial experiences from an ongoing trial deployment in southern Ghana.

**ACM Classification Keywords**
H5.m. Information interfaces and presentation (e.g., HCI): Miscellaneous.

**INTRODUCTION**
There are many kinds of telemedicine, from continuing medical education to patient monitoring [26], but *remote medical consultation* – knowledge-sharing between healthcare workers, focused on specific cases – is the variant most frequently proposed to improve access to specialist expertise in areas where adequate numbers of specialists are simply unavailable. For example, remote consultation projects are often set up to link doctors in developing countries with specialists in North America, Europe and India. Typical projects, funded by research grants or international development agencies, naturally place their primary focus on questions of technological capability or of clinical effectiveness. With a few notable exceptions, most do not address the question of creating stable, self-sustaining communities that will enable the project to continue after external funding ends.

Our research focuses on the fundamental sustainability issue of how to encourage participation and trust in a dynamic, self-organizing telemedicine project in developing regions. We are grounding our work in detailed examinations of three key types of "networks" – social networks of the individual doctors, organizational networks of the medical and governmental institutions, and communication network infrastructure – in the West African nation of Ghana. Ghana is a particularly apt case study due to its extreme disparities in specialist distribution, both at the international level (by conservative measures, 30% of Ghana-trained physicians have migrated overseas [19]) and regional level (specialist knowledge is concentrated in the urban centers). Telemedicine has the potential to bring specialist knowledge to where it is most needed, but only inasmuch as it can be adapted to the local context and conditions.

The contribution of this paper is twofold. First, drawing on two rounds of extended design fieldwork in Ghana interleaved with iterative design, we provide three principles to inform the design of telemedicine systems in developing regions. These principles focus on the use of personal social networks as a framework on which to build incentives within a self-organizing network; optional integration of existing referral mechanisms; and a weakly-connected, distributed architecture that allows for a highly interactive, responsive system despite failures in connectivity. Second, we describe a system that applies these principles and provide initial results from an ongoing trial deployment in Ghana. The system enables doctors to enter medical case information into a distributed repository and then assign the case to (i.e., request a consultation from) members of a pool of medical specialists using a social networking interface. Doctors can also initiate open-ended discussions with groups or individuals. We expect these principles and initial findings will be of interest to designers of knowledge-sharing and telemedicine systems in general and those working in developing regions in particular.

The paper is organized as follows. We first provide an overview of previous work relevant to remote consultation, focusing on research applicable to developing regions. After outlining the fieldwork and iterative design process used in this project, we present an overview of high-level themes arising from this work. After describing the three design principles and their specific grounding in our findings, we give a preliminary discussion how they have fared in our trial deployment. We conclude with suggestions for future research.

**RELATED WORK**
Our ideas build upon decades of work in telemedicine [11,26]. We depart from prior telemedicine work in drawing upon lines of HCI/CSCW research that telemedicine has largely ignored, especially that concerned with the dynamics of latent and articulated social networks.

Telemedicine includes many disparate applications [26], including doctor-to-patient and doctor-to-practitioner tele-clinics (e.g., [2]), distance learning for initial (e.g., [10]) and continuing (e.g., [9]) medical education, patient tele-monitoring and home care (e.g., [1]), and doctor-to-doctor remote consultation (e.g., [13]); consequently, we focus here on remote specialist consultation using asynchronous computer-mediated communication. This is the approach typically applied in developing regions [23] and other locales where transportation infrastructure may be limited and network infrastructure capable of sustaining real-time media connections is not cost-effective [11]. From a technical viewpoint, asynchronous remote consultation systems can be divided into message-, storage-and discussion-centric systems. Message-centric systems provide email(-like) functionality (e.g., [23]), enabling doctors to send questions to specialist consultant groups and receive replies. They are relatively easy to deploy but lack many of the kind of content management features that are useful in establishing communities. Storage-centric systems, such as the Web- or message-based picture archive communication systems (PACS) often used in teleradiology (e.g., [12]), add basic search and storage capabilities. Discussion-centric systems implement the functionality of a typical Web-based bulletin-board system (including messaging, discussions, image storage, etc.) in a package customized for telemedicine. iPath is a notable example of the discussion group model, though it recently added an explicit case assignment system known as the "virtual institute" [4]. Our system builds upon iPath as software substrate, adding new interaction metaphors and storage capabilities described later.

Healthcare systems research outside of HCI/CSCW rarely draws on relevant HCI/CSCW research and detailed design fieldwork or workplace studies are seldom used [21,24]. There are certainly relevant examples in the HCI/CSCW literature: for example, workplace studies in hospitals have described the relevance of individual relationships (e.g., in multi-site X-ray reading [12]) and social networks (e.g., in remote consultations [13]) in distributed medical work, and healthcare projects in the HCI/CSCW community often demonstrate the benefits of user-centered processes (e.g., [1,2,13]). Our work expands upon all of the work above by (1) explicitly considering the problem of establishing a community that can grow in an organic and sustainable way (previous remote consultation systems, including [13], assume pre-existing specialist groups or institutional collaborations) and (2) developing these ideas into primary elements of interface design (social networking features, described later). In this, we draw inspiration from recent research that attempts to understand the dynamics of public social networking sites such as Friendster [3] and Facebook [8] as well as those of health-oriented online communities (e.g., [6,16]).

We also extend telemedicine research by building on HCI/CSCW work in expert-finding. These "second-generation knowledge sharing systems" [22] emphasize the importance of informal knowledge and communication flows in knowledge-sharing system [5]. The mainstream of this research has emphasized the automated discovery of social networks by mining repositories of user data and user interaction [14,17,18,20], using relationships as an input for expert-recommender algorithms; however, now that use of social networking software is relatively common worldwide, expecting users to enter social networks is far less problematic and we follow systems such as ContactMap [25] in directly exposing ego-centric social network data to end-users as they look for experts.

**METHOD**
In this section, we present an overview of our needs assessment, iterative design process, and pilot deployment.

**Needs Assessment Fieldwork**
The initial needs assessment was conducted in Ghana over a five-week period. This phase consisted of site visits to nine medical institutions throughout Ghana (2 regional hospitals in the rural north, 2 major urban hospitals, 5 clinics and polyclinics in the south). Site visits included partnership-building, site surveys, participant observation of two doctors' workdays, and semi-structured interviews (4 rural doctors, 7 urban doctors, 4 support staff). Subjects were recruited by snowball sampling from our existing contacts. We also collected data on the information and communication technology (ICT) infrastructure and computer literacy of the physicians as indicated by the prevalence of word processors, digital cameras, and email usage. Interviews focused on the structure of Ghana's medical system and areas in which ICT might be used to improve it, as well as on medical records, resource limitations, patient diagnosis, and perceptions of technology. At the close of this fieldwork, a focus group was held with five doctors to validate conclusions and envision solutions. Back in the U.S., we applied coding, clustering, and categorization techniques from grounded theory to our data, and identified key themes which will be addressed in the Design Fieldwork section.

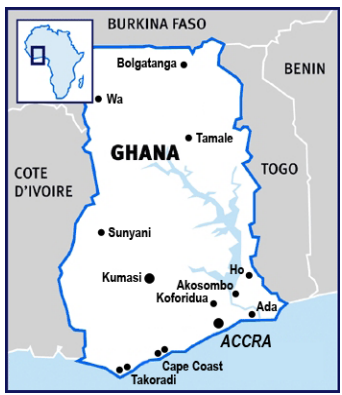

**Figure 1. Ghana fieldwork sites.**

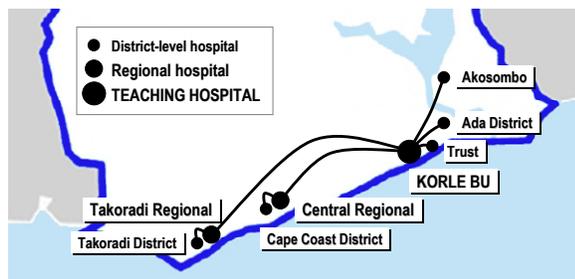

**Figure 2. Referral hierarchy for hospitals in trial deployment.**

### Iterative Design

Based on the needs assessment and a review of relevant telemedicine literature [15], we narrowed our focus to asynchronous remote consultation and began our first set of design iterations. Over a four-month period, we worked with 10 U.S.-resident doctors from a partner organization (seven of whom were from Ghana), conducting semi-structured interviews and assessing paper prototypes.

We returned to Ghana for a second set of design iterations and design fieldwork. We followed the same process as above for an additional five weeks, splitting into two teams to cover the north and south of Ghana (sites visited to date can be seen in Figure 1). In addition, in preparation for a pilot deployment, we conducted focus groups and 20-point site surveys of the IT infrastructure. This time, we visited 18 sites (7 regional hospitals, 8 district hospitals, 3 clinics) and conducted 60 interviews (25 regional hospital doctors, 25 district hospital doctors, 2 nurses, 8 support staff). Participants were chosen with the assistance of each hospital's medical director to provide coverage of junior and senior personnel, general and specialist physicians, and novice and expert computer users.

### Deployment

After completing an initial prototype, we returned to Ghana for a third time, spending two months on an exploratory pilot deployment. First, we hosted two servers with Internet service providers in the U.S. and in Ghana. These enabled participants with any form of home, workplace, or public (e.g., cyber-café) Internet connectivity to access the system. Second, we visited eight hospitals in southern Ghana (Figure 2), installing four local servers in two district hospitals and two regional hospitals in Ghana that had Internet connectivity. As described later, these enabled participants who had intermittent workplace Internet connectivity to access the local server. We conducted group user training at each site and one-on-one follow-up training with 18 participants, including all the doctors in the district hospitals (five in total) and all the doctors at the regional hospitals who were available for interviews on the days we visited. In conjunction with training, we conducted preliminary feedback interviews.

In all, we have enrolled 73 participants in Ghana, the U.S., Mali, Nigeria, South Africa, and the U.K. Participants have been recruited through a combination of on-site recruiting at hospitals, presentations at medical professional group meetings, email to medical professional groups, and personal contacts. Participation is voluntary and uncompensated. Access to case data is limited to participants – explicit patient-identification data is not stored, all data carried over the network is encrypted, and accounts must be created by system administrators.

### DESIGN FIELDWORK

In this section, we highlight some of the background issues that came out of our initial design fieldwork. We defer discussion of issues specific to particular design principles to the next section – here, we provide context that motivates the use of telemedicine and that illustrates the design constraints resulting from conditions on the ground.

The government healthcare system in Ghana is tiered, a common approach worldwide. Doctors and specialists are concentrated at the regional hospitals and the two teaching hospitals in Accra and Kumasi; one or two doctors work at each of the 92 district hospitals; and health workers deliver primary care in local polyclinics and health centers. Cases that cannot be handled at a certain tier are referred up this hierarchy, with all referral chains culminating at the teaching hospitals where specialists have the training and resources to carry out more complex procedures.

Although healthcare is a major concern of Ghana's government and Ghana's medical training program is relatively extensive, a combination of economic, personal and political issues result in widespread hospital staff and equipment shortages. This is especially severe in rural regions – the Upper West region, for example, has eleven doctors for a population of over half a million people [7]. Predictably, a recurrent theme in our interviews was that front-line and rural doctors face heavy workloads and a professionally isolated working environment. As previously mentioned, at least 30% of Ghanaian-trained doctors eventually join the growing professional diaspora, migrating in search of better pay and working conditions and this trend is particularly strong for specialists (those with post-graduate training in surgery, pediatrics, internal medicine, etc.) [19]. To supplement the workforce, the Ghanaian government imports doctors from Cuba and Egypt. Expertise shortages featured broadly in our interviews as well.

The existing referral mechanism, while vital, is problematic for all parties. When a serious but non-emergency case cannot be treated locally and is referred up the chain, the doctor typically gives the patient a paper slip with a few notes. The patient must arrange for transportation, often meaning a trip by mass transit over long distance, at relatively high expense, and over bad roads; many low-income patients cannot overcome these barriers and never go, leading one doctor in the rural north to characterize referrals as death sentences. Receiving doctors find that even when patients arrive with referral forms, there is generally not enough information, and the doctor must rebuild the clinical history by talking to the patient. In one interview the doctor made it a point to say that he could not even determine what medications the patient had been taking prior to arrival. While the slip has a place for referral doctors to provide feedback, slips are rarely seen again by the referring doctor.

We were struck by the degree to which doctors had appropriated their personal social and technical networks to work around these issues. To solve problems locally and thereby avoid more referrals, some doctors use personal contacts, calling or emailing friends at other hospitals to seek advice. Should a referral become necessary, they will call a colleague at the referral hospital to notify them to expect a patient and arrange for a bed. However, junior doctors, especially those placed at rural hospitals, have fewer personal contacts at their referral hospital, especially among older doctors with specialist training. Some hospitals address this informally by posting directories of the (personal) mobile numbers of doctors in the region in their telephone exchange. However, many doctors are uncomfortable calling doctors they do not know; they resort instead to "blind" referrals.

The reliability and availability of telecommunications infrastructure are key barriers in coordinating and consulting with colleagues. Landlines are frequently out of service [27] and mobile networks are unreliable, even when both parties are within service coverage areas. During our deployments, it would often take two or three calls to establish a connection. Likewise, broadband Internet infrastructure is unevenly available, unreliable when available, and dependent on hospital budget allocations. Other options are available throughout the country, such as dial-up, satellite, or mobile data plans, but these are generally expensive, unreliable, and/or slow as well.

In summary, Ghanaian medical professionals face many challenges resulting from unequal distribution of resources, particularly those outside the major urban centers. However, they are also resourceful in their appropriation of ICT to gain access to remote expertise (reducing patient referrals) and to smooth inter-hospital patient management. They are also willing to expend resources to try new strategies; personal mobile phones are very common among Ghanaian professionals, and the directors of the two regional hospitals in our deployment subscribed to broadband Internet service as soon as they learned from us that it had become available in their area. Proposed solutions must nevertheless address the realities of unequal distribution and infrastructure limitations.

**DESIGN PRINCIPLES**

From our initial design fieldwork, we distilled three principles that guided the architecture and interaction design of the system. To ground this discussion we first explain the overall system, and how we envisioned its use.

Our system resembles online social networking services (e.g., [3,8]) and supports two types of conversational threads: case *consultations* and *discussions*. Doctors generally log on to one of our Web servers (1) to refer new cases to colleagues or groups, (2) to review responses to their own cases, or (3) to review cases that have been referred to them. They can also start a free-form discussion with any colleague or group. Other functions support selection of other doctors in the system as colleagues, and identification with particular groups. Groups correspond to particular specialties, institutions/hospitals, countries, and medical professional organizations.

The welcome screen displays a doctor's "primary cases" (i.e., those that were created by, or have been referred directly to, the doctor) at the top (Figure 3, (a)) and the doctor's "other cases" (i.e., those that have been referred to any group of which the doctor is a member) at the bottom (Figure 3, (b)). From this screen they can also choose to view consultations or discussions that have been directed to them, create a new one (Figure 3, (c)), or manage their lists of colleagues and group memberships (Figure 3, (d)). Creating a new consultation entails filling out a consultation request form, in which they specify non-identifying information about the patient, clinical history, and the applicable specialization. In the final screen of this process (Figure 4), they are presented with lists of individuals and groups to whom the consultation can be directed. *Referrals* will be directed towards a referral hospital or a particular colleague with whom they have already made arrangements for patient transfer. *Remote consultations* may be directed towards a colleague with a particular specialty or to a group if the doctor does not have a specific colleague in mind. *Discussions* are exchanges of free-form text on any subject and can be directed to any group or individual.

The system will then notify the selected consultant by email or SMS (a text-message to their mobile phone). Likewise, all doctors are notified of updates to their "primary cases," and optionally their "other cases." Upon receipt of this notification, doctors access the system through one of our public Web servers, or through a Web server locally hosted in the hospital. (We have also set up a number of public computers for the doctors to use to access the Web application in the hospitals, in addition to configuring computers for doctors who have their own.)

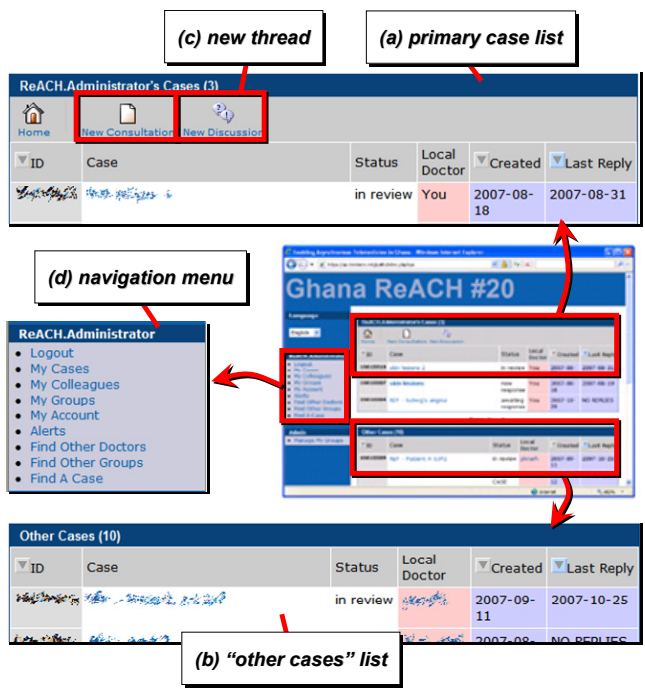

Figure 3. Key elements of the welcome screen.

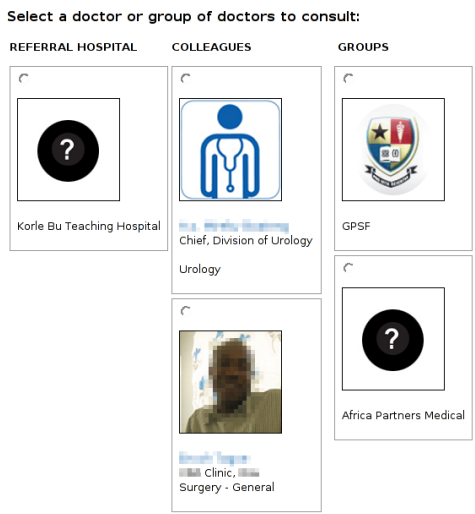

Figure 4. Consultant selection form.

In the remainder of this section, we describe each of the design principles that motivated this organization of our system in turn: using social networking as a primary frame for interaction design, architecting the system to make telemedicine an optional overlay on existing patient-care workflow, and employing what we call split interaction to enable interactive response in the face of unreliable infrastructure.

### Social Networking as a Primary Frame

Social networking is not the primary interaction frame used by existing remote consultation systems. Instead, they work hard to create bridges between and within groups of professional strangers; the iPath system's "virtual institute" [4] mechanism, for example, was designed to improve lagging response times within its pool of consultant doctors by applying the familiar "doctor on call" metaphor. We believe that social networks create an alternative form of accountability that is appropriate for the Ghanaian context. In this section, we examine this assertion more closely. We first unpack the relationship between "social" vs. "professional" and then discuss some specific issues in social networking given the context here.

It should first be understood that medicine, like most professions, is fundamentally a form of group work; workplace studies in hospitals make it clear that there is a strong orientation to colleague relationships and knowledge sharing [12,13]. This was echoed by our participants:

> I always prefer to work in a group. You get more rich knowledge, and you can learn to fit your knowledge to the case.

Once they graduate from the teaching hospitals, doctors posted to the smaller and more distant hospitals often felt professionally isolated. Where once they would gather with other doctors on regular ward rounds, they are now on their own, get little feedback on their practice, and have no one with whom to discuss the range of possible treatments for a given patient. An opportunity clearly exists here for application of mediated communication technologies.

Indeed, as mentioned above, it was clear from our fieldwork that informal use of personal ICT and social networks for remote consultation (as well as referral management) is already common practice. All doctors we spoke to regularly call colleagues to talk about cases.

> I used to work at the main referral hospital before I came here. So I know a number of the doctors who are there. So I usually find out which team is on call, and if know who the doctor is, then I speak to him personally.

In fact, some of our specialist interviews were even interrupted by phone calls from colleagues seeking advice.

While the potential for social networking as a primary frame in Ghana seemed clear from the beginning, it also became clear in the course of the design fieldwork that the system would be significantly different from the archetypal social networking application. In early design iterations, we conceived of the system as a full-featured, open-ended interaction experience along the lines of popular social networking sites [3,8] and online healthcare support communities [6,16]. However, a social networking system for this context involves several key complications.

First, it became clear that the extreme time limitations of the target community meant that the system must be *task-oriented rather than open-ended*. Within the first few interviews it became very clear that doctors would need a reason every time they logged into the system. Once logged into the system, some participants imagined having less-structured discussions and interpersonal relationship building (see the Discussion section). But the design orientation to individual cases would have to be paramount.

Second, as suggested by previous work in accountability in remote consultation [4,15], personal responsibility for answering queries would have to be emphasized. Doctors were clearly already answering queries from their colleagues out of professional and personal courtesy – even the Ghanaian doctors in America had organized themselves into a professional group called the Ghana Physicians and Surgeons Foundation (GPSF) to support their colleagues in Ghana on a purely voluntary basis. But accountability could be expected to vary as the strength of doctor-to-doctor relationships varies; colleague relationships result from various sources (former classmates, many of whom have migrated overseas; colleagues at previous work sites; contacts in training seminars) and many involve additional social ties (social acquaintances; fellow church members; family). The user interface highlights personal assignments.

Third, *assignment to larger groups,* as well as personal contacts, would need to be supported so that the system can leverage its context within existing social structures to ensure initial participation and continuous growth. For example, GPSF actively draws in a community of diasporic doctors, who express a great deal of interest in using the system to re-connect with doctors in Ghana, giving them a means of participating in the medical community they left behind. The many who go back yearly for medical outreaches and specialist training can now integrate this system into their outreach program, so that they can monitor cases created during their time in Ghana and mentor handling of cases that arrive after they leave.

Moreover, larger groups give doctors in the system without a strong existing network a mechanism by which they can actively grow it. If a junior or non-native (e.g. Egyptian or Cuban) doctor does not have a colleague with the right expertise, it would be necessary for them to solve the immediate problem at hand by submitting cases to specialty groups, professional groups, or referral hospitals.

**Telemedicine as Optional Overlay on Existing Workflow**
In addition to the social networking framing just described, we argue that integration with existing formal communication mechanisms is useful and complementary. Thus, for example, doctors in our system are also given the opportunity to request consultations from specialists from the appropriate specialist department in their assigned referral hospital. This may seem counter-intuitive given the emphasis we have placed so far on the importance of personal connections and shortages of expertise within the public healthcare system. In this subsection, we explain how this design decision is equally grounded in doctors' practices and mental schemata for handling these patients that were observed during our design fieldwork.

When we asked our Ghanaian participants to talk about their existing communication patterns with doctors at other hospitals, nearly all of them also talked about the physical referral process. This mental linkage reflects the fact that cases requiring consultation with other doctors are generally difficult ones that are more likely to require physical referral. In other words, the transfer of patients between hospitals, along with the accompanying paperwork and follow-up, is presently the most likely reason for a doctor to communicate with doctors at another hospital.

Our design insight is that the information needed for the referral workflow and the remote consultation workflow are similar enough to be combined into a single set of online forms, enabling the two workflows to be combined up to the point of deciding which doctor(s) will be assigned to the case. In the system, users enter the case data and are then presented with side-by-side listings of both individual contacts (derived from their personal social networks) and departmental contacts (derived from the referral hierarchy of their hospital); assigning the consultation is the final step in submitting the case. Hence, doctors can consult on medical cases with doctors at the referral hospital or elsewhere, or they can send a case directly to the referral hospital department. The primary physician can easily escalate a case entered as a consultation to a referral.

This combined formulation has several advantages. First, it makes the system dual-purpose – it can be used for both remote consultation and referral, which simplifies escalation as well. Second, it makes the end-user's work of requesting a consultation (a "new" task) parallel to that of referral (a task with which doctors are already familiar and consider part of their daily workload). Fitting existing task models should ease training and adoption. Third, it defers the decision of whether to consult remotely or to refer the patient until the final step of the workflow, ensuring a full pass over the case data before the decision is made. Fourth, it allows the system to function as a kind of directory that doctors can use as a convenience and to bootstrap their social networks. Doctors with a weak network can tap into the professional groups in the system, whereas doctors who are newly-assigned to a hospital can use the system's departmental contacts until they can extend their professional networks, whether through the system (when doctors respond) or in person (over time).

Another critical aspect of tapping into the institutional infrastructure is that it gives the system a baseline for trust and professionalism. Although the system is an open network, *institutional gatekeepers* ensure that only certified doctors participate. Interviews indicate this is satisfactory for the majority of cases, although a few specialized cases require the personal recommendation of a specialist in the field.

Even though we have made the referral mechanism available, we believe it is critical that the system remain an optional, low-overhead means of complementing, not replacing, existing referral mechanisms. It must be optional because the primary physicians can run into severe time constraints while handling patients and because the power and network connectivity infrastructure can be unreliable; hospital staff must always have the option to fall back on paper. In addition, as electronic data entry is somewhat slower than paper forms and most day-to-day cases will require neither consultation nor referral, we do not advocate the wholesale use of the system as an electronic medical record (EMR) system.

**Split Interaction: Locally Synchronous, Globally Asynchronous**
As mentioned in the previous section, interaction design must reflect the realities of connectivity and access on the ground. The key idea here is to place the interactive system as close to the user as possible, ensuring a complete user experience with the system even when connectivity is unavailable.

A first design constraint is the need to enable access from both the open Internet and the hospital. Doctors were universally pressed for time while at the hospital, and the four doctors we met who did have computers at a hospital with connectivity tended to use it only at the end of the workday. In addition, all the doctors we asked agreed that many consultations would not require physically being in the hospital, pointing to the need for access from home or from a cyber-café. However, both specialists and general practitioners agreed on the importance of being able to submit images, laboratory findings, and other data which might only be available in-hospital. This dual usage pattern also seems supported by the current and anticipated Internet usage patterns of these physicians: about a third of the physicians interviewed access email regularly from Internet cafes, their laptop PCs, or desktop PCs at home, whereas another third claimed that they had at some point accessed email regularly (and would do so again if it could be accessed more conveniently than by going to a local cyber-café, i.e., at work).

A second design constraint is the need to accommodate existing unreliable infrastructure. Some telemedicine projects install dedicated network links to support high quality-of-service between sites, but these are extremely expensive in Ghana (especially if international links are involved). To remain compatible with our goal of developing an organic, scalable and inclusive network, we must employ the (often unreliable) widely available networking services. Frequent power outages add to the reliability problems.

We examined many possible architectures, including an asynchronous message-based system, a browser-based Web server system, and an asynchronous system combining a frontend PC application with a backend Web server. The email-based telemedicine systems we described previously as being popular in developing regions are suitable to the infrastructure, but clearly insufficient to organize and re-use referral information or to manage and extend one's personal network for consultation. Given the unreliability of power and connectivity, and the fact that fewer than half of the doctors now check their Web-based email regularly, it was clear that a fully-online Web-based system would not be sufficient. At the same time, requiring local installation of a PC application would prevent doctors from freely using the diverse ICT resources available to them (centrally-managed hospital PCs, personal laptops, cyber-cafés). Moreover, a locally-installed application is subject to the vagaries of its host PC (outdated operating systems, unreliable second-hand components, multiple viruses) and is far more difficult to debug, upgrade and evolve.

Our technical solution was to set up a two-tier network. First, we placed highly-available servers on the open Internet, one in the U.S. and one in Ghana. These served as the access points for the doctors outside of Ghana and for Ghanaians with reliable Internet connectivity at home or work. Second, in hospitals with unreliable broadband, we deployed low-cost PC servers to enable users on the local area network to experience a fast connection to the interface even when Internet access is down. Third, we constructed the underlying software so that each of the various servers transmits and globally synchronizes its local database as connectivity becomes available. The system handles this automatically, which required developing application data semantics that can handle asynchronous updates by avoiding or resolving data conflicts cleanly.

As Web designers are well aware, "push" notifications become critical when asynchrony is present but interaction is time-sensitive. An ophthalmologist who had been a participant in a previous telemedicine project impressed upon us the frustrations of using such an "amorphous" system, in which you never know when updates would be available and must regularly log in just to check. And indeed, at present, many participants do not regularly access a computer and need something to draw them into the system. To that end, we set up email and SMS updates which are configured at the local server and issued when new data becomes available.

One potential issue with such a design is that at any given time there may be different snapshots of the data at different servers, which may be confusing to users if they should choose to access the system from both sites. In our discussion, we outline some of the responses we received during our pilot deployment and steps that can be taken to minimize confusion.

## DISCUSSION

In this section, we discuss preliminary findings from our ongoing pilot deployment. The goal of the pilot deployment has been purely exploratory – to test the system in both technical terms (e.g., to uncover unanticipated installation obstacles and software bugs) and conceptual terms (e.g., would users, ranging from medical interns to senior administrators, understand the social networking framing and feel it was appropriate for a medical application?). We begin by discussing our experiences deploying a system in Ghana on a technical level. We then detail our findings about the usage patterns within this medical community, in particular, the high preference for user-to-user interaction. Finally, we touch on the disparity of social connectedness of some groups in the existing system, for whom mechanisms must be made available expand their resources.

Several caveats are in order. First, due to the preliminary nature of our pilot deployment, the discussion draws on only five weeks of operational data, so numbers given (such as the usage patterns) cannot be considered to be more than suggestive as they are subject to novelty effects. Second, the scale is not large. As previously mentioned, 73 participants from various countries were enrolled, 29 of whom have entered social network data into the system (including the 18 in Ghana who made themselves available for individual training).

### Deploying a Split-Interaction System in Ghana

Our decision to build a Web-asynchronous system was validated in two ways. First, we have had been largely successful in troubleshooting and debugging the remote servers and keeping them running even when deployed into organizations with limited technical expertise. For example, when users at one of the district hospitals reported difficulty accessing the system, we were able to log in remotely, check the database, and fix the offending code while maintaining otherwise seamless operation for the user. Another time the server became unavailable due to internal connectivity problems, and even though there is no permanent technical staff at that hospital they were able to identify that it was a local problem and correct the matter with their ISP accordingly. Second, with regards to our assumptions about connectivity, our system has handled frequent outages. Our network monitoring results corroborate our observations about intermittent connectivity: the broadband-connected hospitals do indeed experience regular outages that last from 1 to 4 minutes with great regularity throughout the day, as well as longer outages lasting up to 1.5 hours a few times a week.

In the weeks the system has been running, we have learned two lessons about designing asynchronously-updated systems. First, we did not explicitly consider longer outages due to human interference in our interaction design. For example, the Cape Coast system was unavailable for a full week due to internal network maintenance. Although this is not a regular occurrence, it would be useful to ensure the system notifies the user (in the user interface and by "push" notification) if it has not been able to synchronize over a certain length of time so that participants can adjust their usage accordingly. Second, and more importantly, we did not anticipate the degree of offline or "out of band" collaboration between users in different hospitals. Such communication directly exposes the underlying asynchrony when it is extended by prolonged outages like those just mentioned. One user of the system raised multiple complaints about the fact that his friends from different hospitals (with whom he communicated via cell phone) were seeing cases that were not yet available in his hospital. This is an interesting indicator of the breadth of this doctor's social network, but does result in some confusion about the state of data available in the system. In the future it would be useful to send notifications between servers over the mobile phone network so that they will be able to reflect some minimum information about the global state of data in the system, i.e. so that local servers can show that a certain case has been entered in the system even if the bulk of the case data has not propagated fully.

### The Ghanaian Medical Community

A first glance at the conversational threads in the system gives two immediate indicators that social networking holds promise. The first is that users prefer interacting with other users as much as with groups, even for very hard cases, and the second is that many users at the bottom of the referral chain are considered a valuable, if informal, resource at the top of the referral chain.

Table 1 illustrates how 10 out of 16 conversational threads were directed to specific individuals, even when an appropriate specialist group was available. Many were social discussions that did not deal with a specific medical question, but of the professional threads, 5 out of 11 were directed to individuals. Even users who had previously stated that they would want to post a case to a public forum ended up directing their case to a particular individual. During our training visits, we asked participants about this discrepancy in behavior. Sample reasons include but are not limited to: junior doctors finding their (junior) peers or former advisors/mentors more forgiving of obvious questions; doctors preferring to consult colleagues at nearby hospitals to increase the chances of a face-to-face follow-up; and doctors identifying specific individuals as nationally recognized, uniquely qualified experts on a given topic. Most reasons related back to our fieldwork findings on the importance of doctors' personal social networks, with expedience being an important factor as well. This shows promise, since building a system focused on user as well as group interaction is the core of social networking.

| Specialization Consulted | Professional | Social | Total |
|---|---|---|---|
| None | 1 | 4 | 5 |
| Internal Medicine | 4 | 1 | 5 |
| Urology | 2 | 0 | 2 |
| Pediatrics | 1 | 0 | 1 |
| Ob/Gyn | 1 | 0 | 1 |
| Surgery | 1 | 0 | 1 |
| Opthamology | 1 | 0 | 1 |

| Individual vs. Group | Professional | Social | Total |
|---|---|---|---|
| Individual | 5 | 5 | 10 |
| Group | 6 | 0 | 6 |

| Overseas vs. Local | Professional | Social | Total |
|---|---|---|---|
| Overseas Individual/Group | 4 | 1 | 5 |
| Local Individual/Group | 4 | 4 | 8 |
| Worldwide Group | 3 | 0 | 3 |

**Table 1. Categorization of analyzed conversational threads.**

A potential benefit of our approach is the ability for doctors to draw from resources in unexpected places. In Figure 5, consider nodes 20 and 34 (doctors at district hospitals) and node 11 (a doctor at a private hospital not associated with the public hospital network). These doctors are nominally at the "bottom" of the referral chain yet their connections bridge subgraphs that would otherwise be distantly connected or unconnected. This unexpected availability of expertise is not reflected in the formal referral hierarchy, yet can be tapped through social networking.

One unanticipated phenomenon was the number of doctors who chose to consult with overseas doctors of a particular specialty *even if they had had no previous interaction with that doctor.* Fully half of the medical cases submitted to individuals were directed to "any [appropriate specialist] in the U.S. or the U.K.," in the apparent belief that doctors from these countries must have expertise lacking in Ghana. Physicians generally acknowledge that resources and equipment overseas are better, leading to better opportunities for medical education and doctors with different levels of expertise. While this cannot substitute for local expertise regarding local epidemiology, the ability to draw on international connections is clearly desired.

**Creating a Context for New Connections**
Our prior expectation was that the existing professional framework and other groups would provide opportunities to expand weak social networks. The observed behavior of poorly-connected individuals, such as (1) junior doctors, (2) doctors from Cuba, and (3) doctors in the district hospitals, suggested that they did rely on formal or institutional groups. In the case of junior doctors, five out of seven active within the system (e.g., nodes 38, 42, and 62 in Figure 5) had few colleagues and knew few specialists; all of the non-social consultations/discussions submitted by junior doctors were submitted to specialist groups rather than to individuals. In the case of Cuban doctors (e.g., nodes 48, 55, and 58), poor inter-hospital connections are also visible; of the two cases submitted by Cubans in the system, one was assigned to another Cuban and the other was assigned to a doctor in the same hospital. The final case, doctors from district hospitals, was the most interesting. Doctors from the district hospitals are clearly more isolated than those in the (larger) regional hospitals, and although we only had two district hospitals with about two doctors each in our pilot, participants from multiple locations reiterated the importance of the referral hierarchy for these doctors. For instance, a regional hospital doctor commented that district hospital doctors would feel a greater need to communicate along the referral chains – it is more common to refer from district to regional hospitals than from regional to teaching hospitals (Accra or Kumasi). This is consistent with the fact that three out of the four district doctors who submitted a case submitted to a group.

Although we have not been able to systematically test the utility of the referral hierarchy encoded in this telemedicine network, the fact that disconnected doctors draw from these organizational frameworks is a promising indicator that this framework may turn out to be useful scaffolding for junior, foreign-trained, or physically-isolated doctors.

**CONCLUSION**
In this paper, we have demonstrated the utility of an integrated social networking approach to remote consultation for the Ghanaian medical community and diaspora. Based on the current formal and informal practices for referral between hospitals and the highly connected nature of this medical community, both at home and abroad, we argue that a system can be designed for dual use in consultation and referral. As is the case for any "development" project, it is important to keep in mind the haves and have-nots – to provide professional and institutional scaffolding as a resource for those who are not already well-connected. However, the use of such a system must overlay, not replace, existing practice, simply because power outages and Internet failures are frequent and cannot bring the workflow to a halt. Further, the design must be sensitive to the needs of users who are pressed for time, even if connectivity is temporarily unavailable. To that end, we also propose a locally synchronous, globally asynchronous architecture which has potential for a variety of interactive, distributed systems in the developing world.

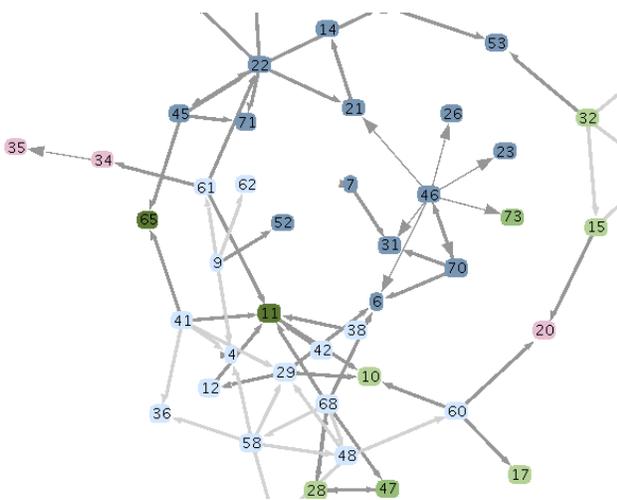

**Figure 5. Visualization of the colleague network.** Arrows point from a participant to users that the participant lists as a colleague. Lighter edges are within-hospital connections while darker edges are between-hospital. Different hospitals are different colors.

The system deployment is ongoing and we expect to continue collecting and analyzing data, in particular with regards to the clinical effectiveness of our system, as well as adding functionality and reaching out to more district hospitals. Given the diversity of the current system use, we look forward to the usage patterns that arise within different communities in this system.

**ACKNOWLEDGEMENTS**
This material is based in part upon work supported by the National Science Foundation under Grant No. 0326582.